\documentclass{article}

\usepackage{arxiv}

\usepackage[utf8]{inputenc} 
\usepackage[T1]{fontenc}    
\usepackage{hyperref}       
\usepackage{url}            
\usepackage{booktabs}       
\usepackage{amsfonts}       
\usepackage{nicefrac}       
\usepackage{microtype}     
\usepackage{graphicx}
\usepackage{doi}
\usepackage{amsmath}

\usepackage[style=numeric,sorting=none]{biblatex} 
\title{Binaspect - A Python Library for Binaural Audio Analysis, Visualization \& Feature Generation}

\author{Dan Barry\\
	School of Computer Science\\
	University College Dublin\\
	Dublin, Ireland\\
	\texttt{danbarry@duck.com} \\	
    \And
	Davoud Shariat Panah\\
	School of Computer Science\\
	University College Dublin\\
	Dublin, Ireland\\
	\texttt{davoud.shariatpanah@ucd.ie} \\
    \And
	Alessandro Ragano\\
	School of Computer Science\\
	University College Dublin\\
	Dublin, Ireland\\
	\texttt{alessandro.ragano@ucd.ie} \\
    \And
	Jan Skoglund\\
	Google LLC\\
        San Francisco, CA\\
	\texttt{jks@google.com} \\
    \And
	Andrew Hines\\
	School of Computer Science\\
	University College Dublin\\
	Dublin, Ireland\\
	\texttt{andrew.hines@ucd.ie} \\
}

\date{}

\renewcommand{\shorttitle}{Binamix}

\hypersetup{
}

\begin{document}
\maketitle

\begin{abstract}
We present \textit{Binaspect}, an open-source Python library for binaural audio analysis, visualization, and feature generation. Binaspect generates interpretable “azimuth maps” by calculating modified interaural time and level difference spectrograms, and clustering those time-frequency (TF) bins into stable time-azimuth histogram representations. This allows multiple active sources to appear as distinct azimuthal clusters, while degradations manifest as broadened, diffused, or shifted distributions. Crucially, Binaspect operates blindly on audio, requiring no prior knowledge of head models. These visualizations enable researchers and engineers to observe how binaural cues are degraded by codec and renderer design choices, among other downstream processes. We demonstrate the tool on bitrate ladders, ambisonic rendering, and VBAP source positioning, where degradations are clearly revealed. In addition to their diagnostic value, the proposed representations can be exported as structured features suitable for training machine learning models in quality prediction, spatial audio classification, and other binaural tasks. Binaspect is released under an open-source license with full reproducibility scripts at \href{https://github.com/QxLabIreland/Binaspect}{https://github.com/QxLabIreland/Binaspect}.

\end{abstract}



\section{Introduction}
\label{sec:intro}

Spatial audio has experienced a renaissance in recent years, with applications ranging from virtual reality to immersive media, as well as the auditory science that supports it. Emerging audio formats such as Dolby Atmos \cite{DolbyAtmos} and Google Eclipsa \cite{iamf} are establishing the platforms on which today’s spatial audio ecosystem is being developed. In parallel, the substantial increase in headphone-based media consumption over the past decade \cite{rane2022survey} has accelerated improvements in binaural rendering techniques needed to deliver spatial audio through headphones. Furthermore, as research into renderer design \cite{rudzki2025eclipsa} and objective evaluation metrics for spatial audio \cite{ambiqual}\cite{panah2025binaqual} advances, there is a need for visualization and analysis tools to help inform design decisions in binaural technology.  

Some useful functionality for binaural analysis is scattered across various MATLAB and Python libraries \cite{soendergaard2013amt}\cite{scheibler2018pyroomacoustics}\cite{majdak2013sofa}\cite{yang2021torchaudio}, but there is no dedicated Python library for binaural feature visualization and generation that the authors are aware of.

Furthermore, we draw on concepts from the world of sound source separation, which we have found to be useful in visualizing binaural signals. In \cite{rickard2000duet}, the authors used interaural time delay (ITD) and interaural level differences (ILD) between closely spaced microphones to cluster and separate TF bins according to their direction of arrival (DOA). They produced intuitive 2D histograms for each time frame, which clearly showed source activity as it related to DOA. Later work modified this approach to operate more efficiently on intensity panned stereo recordings \cite{barry2004azimuth} using a 1D histogram to display pan position and intensity of multiple sources in real time. This approach used interaural level ratio (ILR) instead of ILD, which is essentially a linear gain ratio instead of a dB level difference. 
This panning histogram concept was further utilized in \cite{barry2007music} for music structure segmentation, where the time dimension was added and it was referred to as an "azimugram". It clearly visualized the azimuth trajectory of multiple sources over time. The same time-azimuth visualization was used again in \cite{barry2009localization} for the purposes of stereo to surround upmixing. Later, the TF based ITD and ILD features from \cite{rickard2000duet} were utilized in binaural direction estimation \cite{dietz2011auditory} and included in the Auditory Model Toolbox \cite{soendergaard2013amt}, but the histogram representations described above were not utilized and have not yet seen widespread use in binaural analysis. In this paper, we show their utility for a range of binaural inspection tasks and release Binaspect under an open-source license with full reproducibility scripts\footnote{\href{https://github.com/QxLabIreland/Binaspect}{https://github.com/QxLabIreland/Binaspect}}. 

\section{Feature Definitions}
\label{sec:methods}

\begin{figure*}[h!]
    \centering
    \includegraphics[width=0.7\linewidth]{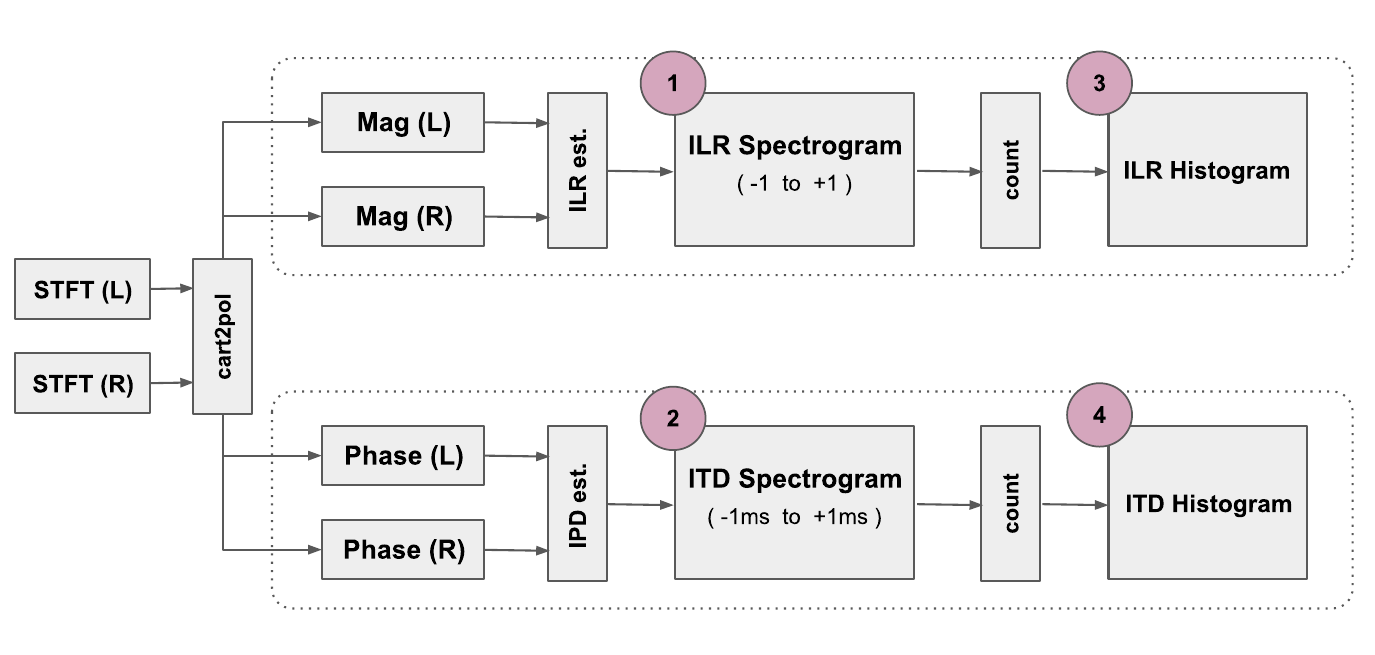}
    \caption{Block diagram illustrating the processing blocks to produce the 4 primary features.}
    \label{fig:block_diagram}
\end{figure*}

A block diagram of the proposed system is shown in Figure \ref{fig:block_diagram} in which we create four representations: (1) Bounded ILR Spectrogram, (2) ITD Spectrogram, (3) Bounded ILR Histogram, and (4) ITD Histogram. The sections that follow explain how to calculate each representation.   

\subsection{Calculating the Bounded ILR Spectrogram}
\label{ssec:ILRspect}
Traditional ILD representations are unbounded and log-scaled \cite{dietz2011auditory}\cite{birchfield}, which can be problematic in machine learning applications. Here, we define a \emph{bounded interaural level ratio} (ILR), which maps raw ILR values into the range \([-1,+1]\) with \(0\) at the center. The bounded \emph{ILR spectrogram} is obtained by first calculating the raw interaural level ratio (ILR) at each time-frequency bin \eqref{eq:ilr_ratio}, and then mapping it into the bounded interval \([-1,1]\) with zero at the center \eqref{eq:ilr_bounded}. The resulting bounded ILR spectrogram provides a stable, signed representation of binaural level cues across time and frequency, where each value can be interpreted directly as a linear gain coefficient. 

Conveniently, the dimension and domain of the bounded ILR spectrogram match that of the magnitude spectrogram and phase spectrogram, so it can be used as an input feature alongside other time-frequency representations in machine learning based source localization systems \cite{phokhinanan2023binaural}\cite{kuang2025bast}.

\begin{subequations}
    \label{eq:ilr}
    \begin{align}
    ILR_{k,m} &= \frac{|R_{k,m}|}{|L_{k,m}|} && \label{eq:ilr_ratio} \\[4pt]
    bILR_{k,m} &=
        \begin{cases}
        ILR_{k,m} - 1, & ILR_{k,m} < 1, \\[4pt]
        1 - \dfrac{1}{ILR_{k,m}}, & ILR_{k,m} \ge 1,
        \end{cases}
    && \label{eq:ilr_bounded}
    \end{align}
\end{subequations}
where \(k\) is the frequency bin index, and \(m\) is the time frame index.

\subsection{Calculating the ITD Spectrogram}
\label{ssec:ITDspect}

Similar to \cite{rickard2000duet} and \cite{dietz2011auditory}, the \emph{ITD spectrogram} is obtained by first computing the interaural phase difference at each time-frequency bin, then wrapping it into the principal interval \([-\pi,\pi]\), and finally converting it to an interaural time delay in seconds. Specifically, the raw phase difference, \(IPD_{k,m}\), is calculated as the difference between the left and right STFT channels \eqref{eq:phase_difference}, the phase difference is wrapped \eqref{eq:phase_wrap}, and the corresponding ITD value is computed by normalizing the wrapped phase difference with respect to the bin frequency \eqref{eq:time_delay}. The result is an ITD spectrogram, \(ITD_{k,m}\), that describes binaural timing cues across time and frequency.

\begin{subequations}
    \begin{align}
         IPD_{k,m} &= \angle R_{k,m} - \angle L_{k,m} && \label{eq:phase_difference} \\[6pt]
         \tilde{IPD}_{k,m} &= \bigl(( IPD_{k,m} + \pi) \bmod 2\pi\bigr) - \pi && \label{eq:phase_wrap} \\[6pt]
         ITD_{k,m} &= \frac{ \tilde{IPD}_{k,m}}{2\pi \cdot k_{\text{width}} \cdot k} && \label{eq:time_delay}
    \end{align}
\end{subequations}

where \(k\) is the frequency bin index, \(m\) is the time frame index, and \(k_{width}\) is the frequency bin width.\\

It is well established that the ITD depends on head geometry and source direction. For a simplified straight ear-to-ear distance of approximately 0.18\,m, the maximum ITD is estimated at approximately 520\,$\mu$s. More accurate spherical head models, such as Woodworth's approximation, account for diffraction around the head and predict lateral-source ITDs of 650-700\,$\mu$s \cite{woodworth1954experimental}. In practice, individualized head shapes and pinna geometry can yield even larger values, and ITDs up to approximately 880\,$\mu$s are observed for extreme lateral sources \cite{Armstrong2018Perceptual}.

As a result of this head geometry which is encoded in the binaural signal, we can only accurately calculate the ITD up to a frequency of approximately 1500~Hz, the wavelength of which corresponds to the average delay across the head. Once the wavelength becomes shorter than the maximum head delay, it becomes difficult to accurately estimate ITD due to periodic phase difference aliasing. This is because for higher frequencies ($> 1500$\,Hz), multiple periods can pass within the maximum time delay between the ears, leading to aliasing errors in the ITD estimates. For this reason, we band-limit the ITD calculations for visualization purposes but still allow full bandwidth processing for the purposes of feature generation in machine learning applications.

Similar to the bounded ILR Spectrogram in Section \ref{ssec:ILRspect}, the ITD spectrogram has the same dimension and domain as other time-frequency representations, making it possible to use it as a feature alongside other TF representations in machine learning applications. 

It should be noted that the ITD and bounded ILR spectrograms are not particularly useful for human inspection, but they are the intermediate representations required to calculate the ITD Histogram and bounded ILR Histogram, which are very useful visualizations for analysing how binaural cues are affected by various types of processing, as shown in Section \ref{sec:results}.

\subsection{Calculating the Bounded ILR Histogram}
\label{ssec:ILRhist}

The \emph{bounded ILR histogram} provides a frame-wise representation of interaural level ratios by accumulating per-bin ILR values into discrete histogram bins. At each time-frequency point, the raw ILR is mapped into the bounded interval \([-1,1]\) according to Equations~\eqref{eq:ilr_ratio}-\eqref{eq:ilr_bounded}, yielding the bounded ILR spectrogram. These values are then assigned to the nearest histogram bin center \(c_b\), and each histogram bin is incremented by a weight \(w_{k,m}\) as defined in Equation~\eqref{eq:bILR_hist_energy}. 
The resulting intensity weighted histogram \(H^{\text{bILR}}_{b,m}\) captures the spatial distribution of sources across time, where stable sources appear as sharp azimuthal clusters, while spatial degradations due to processing (e.g., codec compression) are reflected as broadened, diffused, or shifted clusters.

\begin{equation}
H^{\text{bILR}}_{b,m}
 = \sum_{k}\,
\big(w_{k,m}\big)\;
\mathbf{1}\!\left(\,\big|bILR_{k,m} - c_b\big| \le \tfrac{\Delta}{2}\,\right)
\label{eq:bILR_hist_energy}
\end{equation}

\noindent
Where, \(w_{k,m} = |L_{k,m}| + |R_{k,m}|\), and \(c_b\) is the bin center of the \(b^{\text{th}}\) histogram bin, and \(\Delta\) is the histogram bin width.

\subsection{Calculating the ITD Histogram}
\label{ssec:ITDhist}

Here, we define the \emph{ITD histogram}, which provides a frame-wise summary of interaural time differences by accumulating per-bin ITD estimates into discrete histogram bins. For each time-frequency point, the wrapped phase difference is converted to a delay value \(\mathrm{ITD}_{k,m}\) according to Equations~\eqref{eq:phase_difference}--\eqref{eq:time_delay}. For each time frame, these delay values are assigned to the nearest histogram bin center \(c_b\), and the corresponding bin is incremented by a weight \(w_{k,m}\), which is equal to the magnitude spectrum value at that TF point. Formally, the histogram value is defined in Equation~\eqref{eq:itd_hist_energy}, where \(H^{\text{ITD}}_{b,m}\) represents the energy-weighted count for the \(b^{\text{th}}\) histogram bin at frame \(m\). The result is a time-delay representation in which stable sources appear as sharp ridges in the histogram, while degradations from codecs or rendering introduce broadened or shifted distributions.

\begin{equation}
H^{\text{ITD}}_{b,m}
 = \sum_{k}\,
\big(w_{k,m}\big)\;
\mathbf{1}\!\left(\,\big|\mathrm{ITD}_{k,m} - c_b\big| \le \tfrac{\Delta}{2}\,\right)
\label{eq:itd_hist_energy}
\end{equation}

\noindent
where \(w_{k,m} = |L_{k,m}| + |R_{k,m}|\), and \(c_b\) is the bin center of the \(b\)-th histogram bin, and \(\Delta\) is the histogram bin width.\\ 

Typically, 400 histogram bins provide good visual resolution for human inspection, but other values may be optimal when the feature is used in machine learning applications. Examples in Section \ref{sec:results} show the utility of these histogram visualizations for analysing how binaural cues are affected by various types of processing.

\section{Analysis Examples}
\label{sec:results}
Here, we present three real-world applications where the ITD Histogram and bounded ILR Histogram can be used to analyse how binaural cues are degraded by various processes.

\subsection{Ambisonic Renderer Analysis}
\label{ssec:ambisonic_example}

\begin{figure}[h!]
    \centering
    \includegraphics[width=0.7\linewidth]{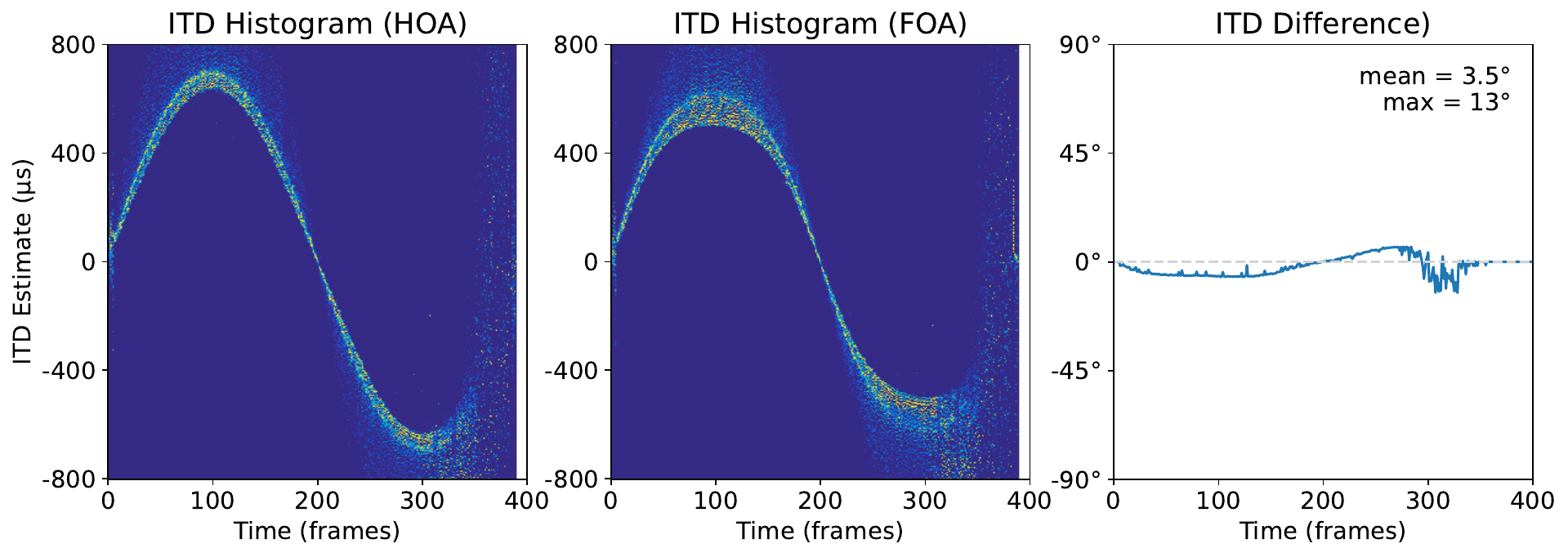}\\[1ex]
    \includegraphics[width=0.7\linewidth]{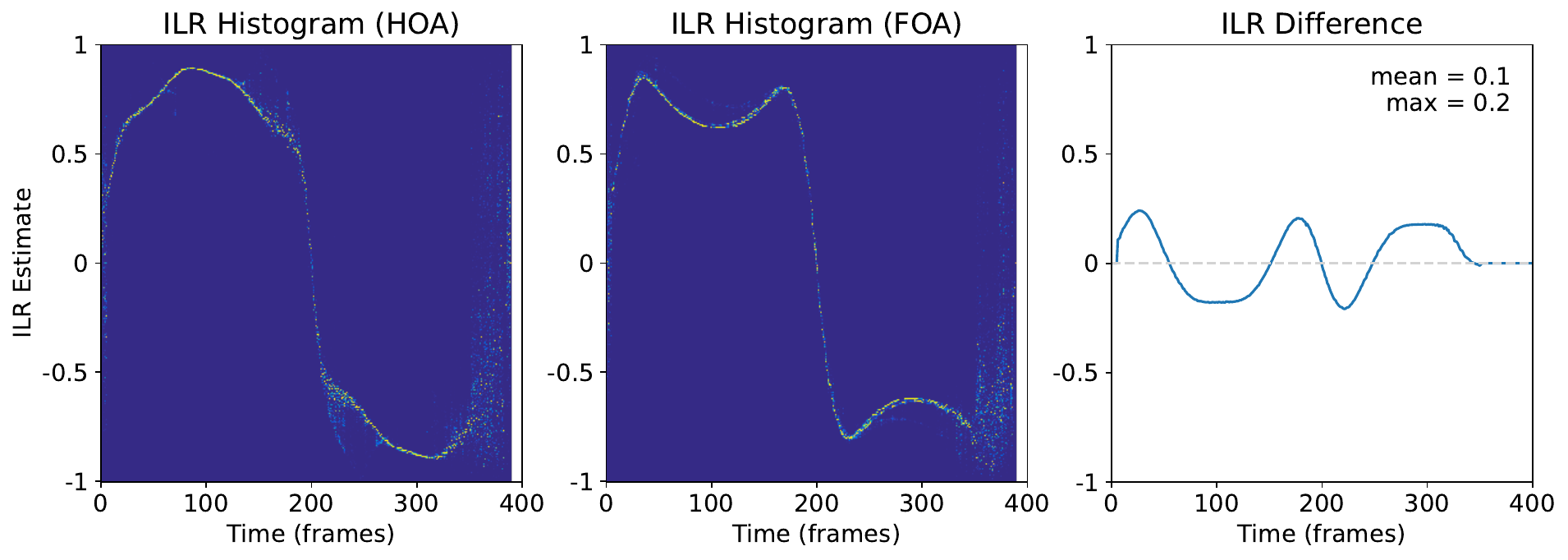}
    \caption{ITD and ILR histograms showing differences between a binaural render of Higher Order Ambisonics (HOA) and First Order Ambisonics (FOA) versions of an audio source being spatially panned from azimuth 0° to 270° with fixed elevation 30°. The respective ITD and ILR differences are also shown.}
    \label{fig:ambi_comparison}
\end{figure}

In Figure \ref{fig:ambi_comparison}, we show the utility of the visualizations to compare binaural renders of a signal encoded in higher order ambisonics (HOA) and first order ambisonics (FOA). We used the Resonance renderer \cite{resonanceaudio_matlab} with subject "D2" from the SADIE II Database \cite{Armstrong2018Perceptual}, which contains a set of high-resolution HRIR and BRIR measurements for spatial audio research. The signal is castanets being spatially panned from azimuth 0° to 270° with a fixed elevation of 30°. The signal duration is approximately 9 seconds. The last 2 seconds of audio are the reverberant tail of the last castanets hit.

Observing the ITD histograms for HOA and FOA, we can see that the source is correctly tracked as it pans around the head. However, the degradation of both ITD and ILR cues is clearly visible in the FOA render. In FOA, the ITDs for lateral sources (\(\pm\) 90°) are notably smaller than HOA. We can also see that the TF points are not as tightly clustered as in HOA. Averaging for each time frame and converting ITD back to degrees (ITD Difference plot), we see an estimated mean angular shift of 3.5° across all TF points. Observing the ILR histograms, we see an even more pronounced degradation in FOA. Here, as the source approaches \(\pm\) 90°, the ILRs begin to decrease notably instead of increasing monotonically as in the HOA case. These binaural degradations will have a notable effect on perception, but for clarity, FOA is inherently limited compared to HOA, so this is not a new finding. The point is to illustrate how useful the visualization can be in diagnosing problems when building render pipelines. In the same way, it is possible to compare different binaural renderers or HRTF sets, for example. The plots shown in Figure \ref{fig:ambi_comparison} are generated by Binaspect using:

{\small
\begin{verbatim}
ITD_diff(ref, test, sr, plots=True)
ILR_diff(ref, test, sr, plots=True)
\end{verbatim}
}

\subsection{Codec Compression Analysis}
\label{ssec:encoder_example}

\begin{figure}[h!]
    \centering
    \includegraphics[width=0.7\linewidth]{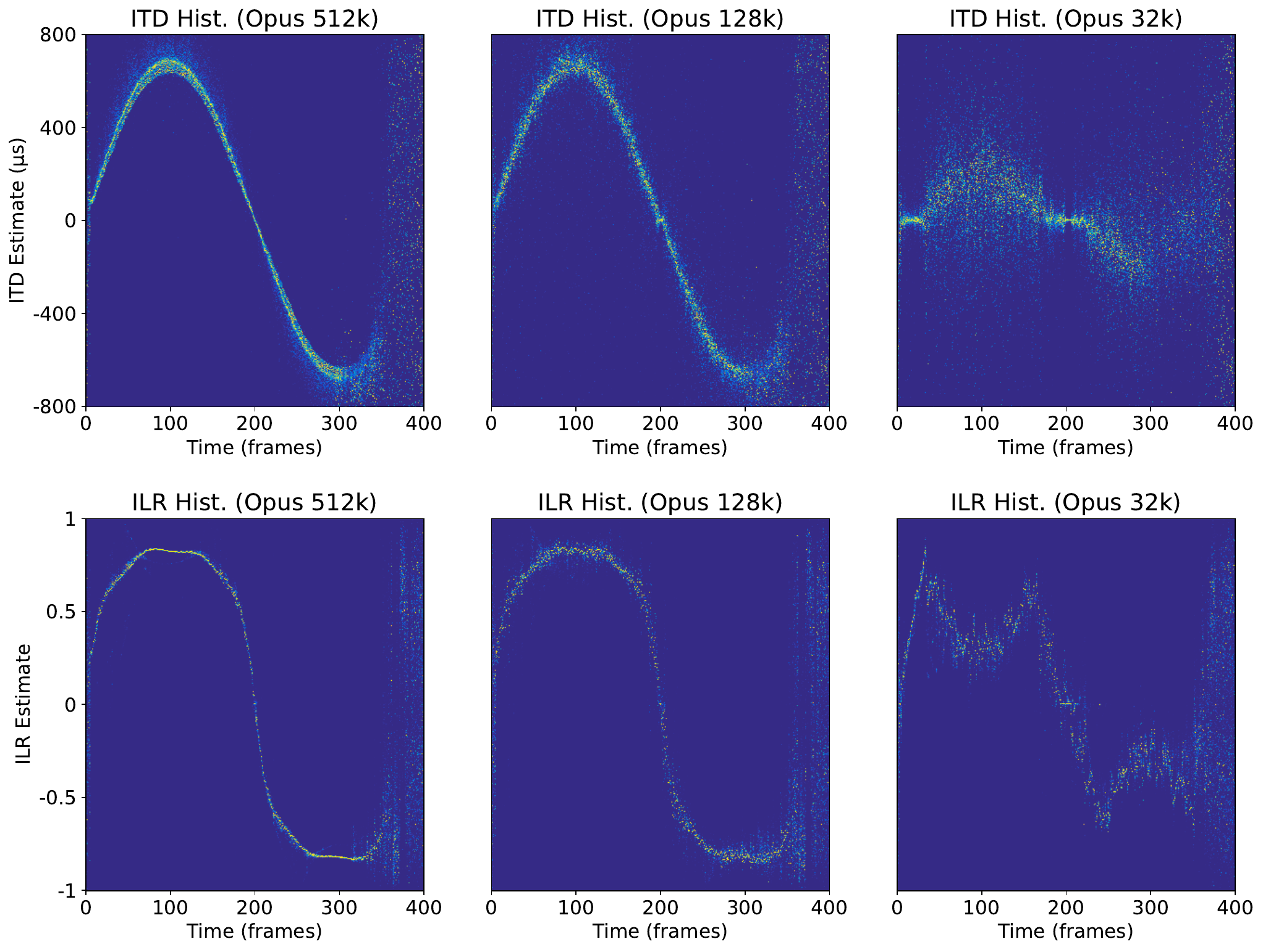}
    
    \caption{ITD and ILR histograms showing differences between various Opus codec bitrates. The audio source is being spatially panned from azimuth 0° to 270° with fixed elevation 30°}
    \label{fig:opus_comparison}
\end{figure}

In Figure \ref{fig:opus_comparison}, we show the utility of the visualizations in comparing the effects of codec bitrate on binaural audio. The signal used here is the same as the HOA example above, except that it is binaurally rendered with subject "D1" from the SADIE II database \cite{Armstrong2018Perceptual}. The rendered signal is then subjected to Opus audio compression at bitrates 512k, 128k, and 32k.

Observing Figure \ref{fig:opus_comparison}, it is clear to see that ITD and ILR TF points remain tightly clustered for 512k and become more diffused for 128k, but still follow expected trajectories. However, the binaural cues for a bitrate of 32k do not exhibit the expected trajectories and are quite scattered and shifted. The perceptual degradation is equally notable on audition. In the same way, it is possible to compare different codecs at equivalent bitrates. Single histograms are generated using Binaspect as follows:

{\small{}
\begin{verbatim}
ITD_hist(input_file, sr)
ILR_hist(input_file, sr)
\end{verbatim}
}

Optional arguments for histogram size, frame normalization, and intensity weighting are also available.

\subsection{Down-mix Analysis}
\label{ssec:downmix_example}

\begin{figure}[h!]
    \centering
    \includegraphics[width=0.7\linewidth]{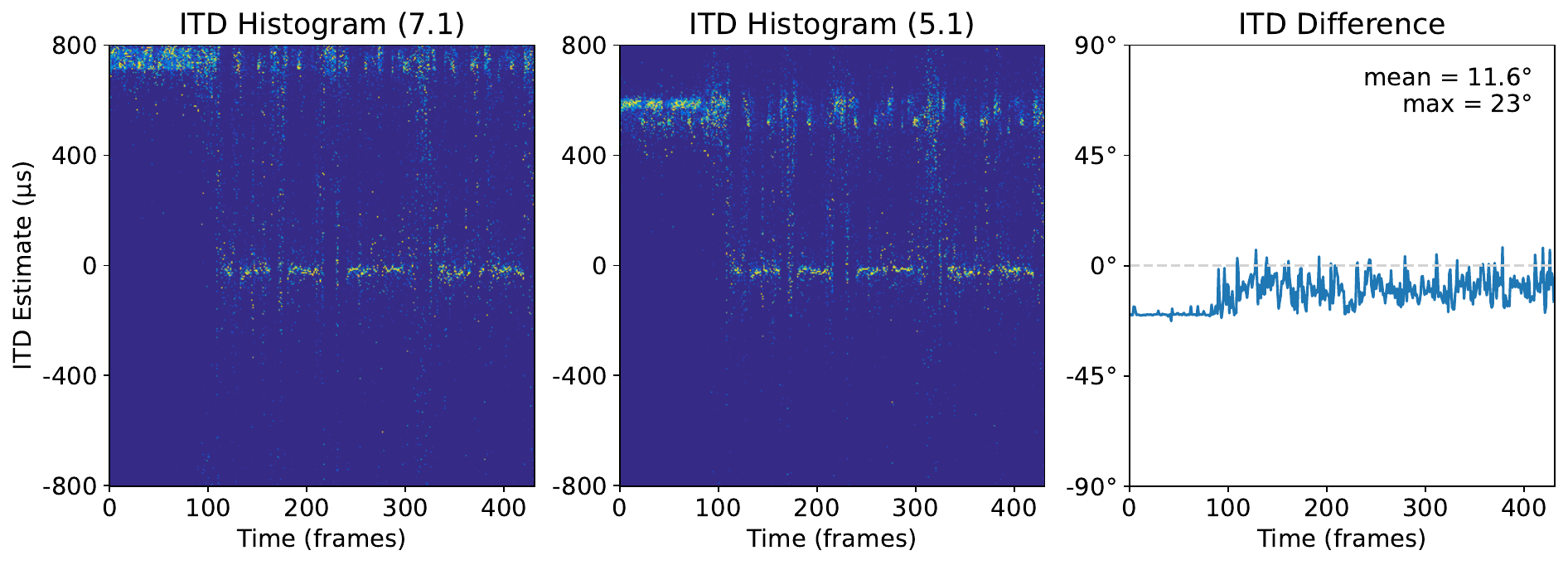}\\[1ex]
    \includegraphics[width=0.7\linewidth]{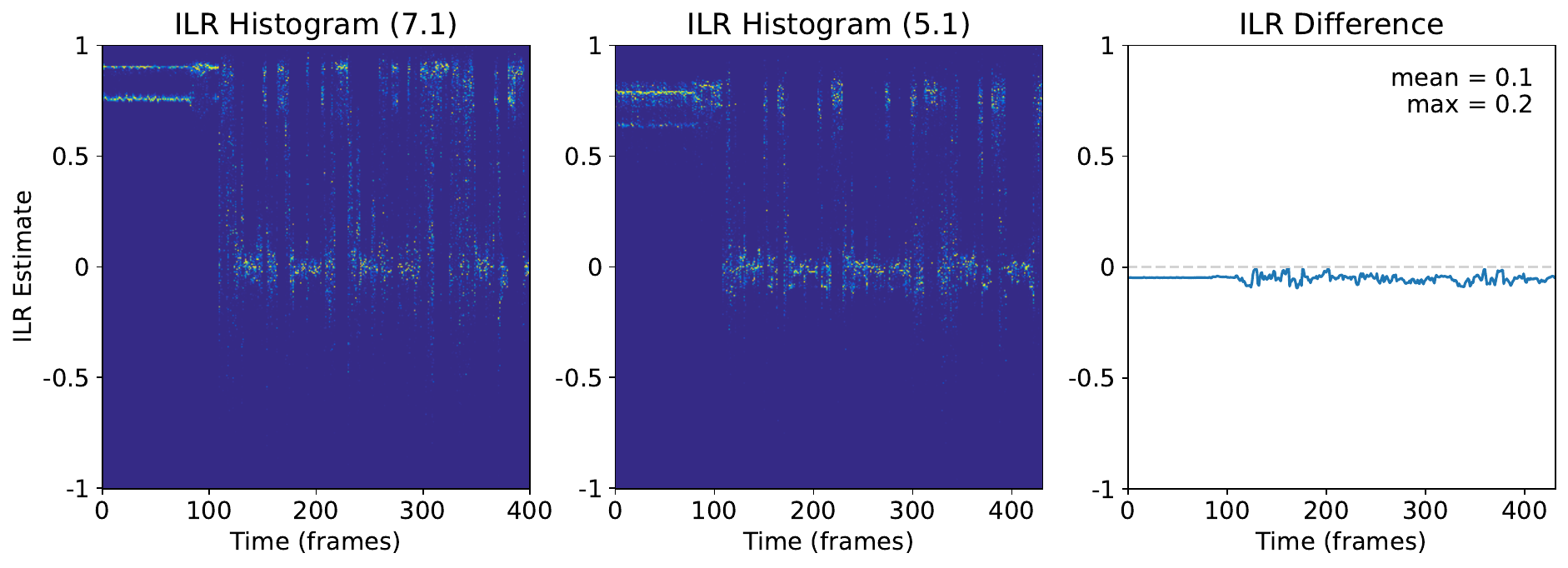}
    \caption{ITD and ILR histograms showing differences between direct binaural renders of a 7.1 mix and a 5.1 mix of the same material. There is one source at 90° and another at 0°. The 90° source has a discrete channel in 7.1 but is rendered as a virtual source in 5.1}
    \label{fig:downmix_comparison}
\end{figure}

In Figure \ref{fig:downmix_comparison}, the histograms show how binaural cues are degraded by virtual source positioning. This happens frequently in down-mixing processes which use vector base amplitude panning (VBAP) \cite{pulkki1997virtual}, for example. The signal used here contains two sources, a vocal track positioned at azimuth 0° and a drum track positioned at 90°. The binaural renders were created using Binamix \cite{Binamix2025} with subject "H6" from the SADIE II database \cite{Armstrong2018Perceptual}. Binamix can be configured to simulate channel-based surround renders by using discrete HRIRs for angles that exist in the chosen channel layout, and VBAP to simulate angles that are not present in that layout. In this example, we compare the mix in 7.1 and 5.1. Both layouts have 0°, but only 7.1 contains a 90° channel. In 5.1, the 90° source must be virtually positioned between the front left (30°) and rear left (120°) channels. 

Observe in Figure \ref{fig:downmix_comparison}, two sources can be clearly seen as straight "tracks" along the time axis. This illustrates the utility of the histogram representations to visualize the DOA of multiple sources simultaneously. The 90° source (drums) appears towards the top of the plot for both 7.1 and 5.1, but in the 5.1 example, it has been notably shifted towards the center in both the ITD and ILR plots. This is because it has been virtually positioned in the 5.1 case, as there is no 90° channel. The ITD and ILR histogram representations clearly show the spatial image shifts and the average shift over time.

Audio demonstrations for all the examples discussed here can be found at:\\ \href{https://golisten.ucd.ie/task/mushra-test/68caa09e4f13034dd15e8653}{https://golisten.ucd.ie/task/mushra-test/68caa09e4f13034dd15e8653}

\section{Discussion and Conclusions}
\label{sec:discussionconclusion}
We have introduced Binaspect, an open-source Python library for binaural audio analysis, visualization, and feature generation. The library provides four complementary representations inspired by early sound source separation research: the bounded ILR spectrogram, ITD spectrogram, bounded ILR histogram, and ITD histogram. While the spectrogram forms are valuable as features for machine learning pipelines, the histogram-based representations are particularly effective for human inspection. They allow stable sources to be tracked as compact clusters in azimuth, while degradations arising from various real-world processes manifest as broadened, shifted, or diffused distributions.

Binaspect complements existing auditory modeling resources such as the AMT by offering intuitive diagnostic visualizations and Python-native feature generation. 
In addition to the ILR and ITD representations presented in this paper, Binaspect also supports many of the traditional binaural representations such as IPD and ILD, with further standard binaural features planned in future releases. 
Every feature provides functions for band limiting and mask generation for use in machine learning. 

The examples presented demonstrate the utility of the approach for analyzing ambisonic rendering, codec compression, and down-mixing artifacts. These case studies highlight both the interpretability of the visualizations and the potential of the exported features for training machine learning models in spatial audio quality assessment and source localization. 

There are some limitations of the system. As the number of sources increases, so too does the amount of TF overlap between the sources. This leads to errors in both ITD and ILR estimates and ultimately makes the histogram representations less useful for visual inspection, but this is a common problem in DOA estimation algorithms.

In conclusion, Binaspect offers researchers and practitioners a practical tool for inspecting, quantifying, and learning from binaural cue degradations across diverse audio processes. Future work will extend the library with additional binaural features, improved handling of multi-source scenarios, and further integration with machine learning frameworks for objective quality modeling and spatial audio classification.

\section{Acknowledgments}
This work was conducted with a research grant from Taighde Éireann – Research Ireland co-funded under the European Regional Development Fund under Grant Numbers 12/RC/2289\_P2 and a gift from Google. For the purpose of Open Access, the author has applied a CC BY public copyright license to any Author Accepted Manuscript version arising from this submission.

\printbibliography

@article{Armstrong2018Perceptual,
	author = {Armstrong, Cal and Thresh, Lewis and Murphy, Damian and Kearney, Gavin},
	journal = {Applied Sciences},
	doi = {10.3390/app8112029},
	issn = {2076-3417},
	number = {11},
	year = {2018},
	month = {11},
	note = {number: 11
publisher: Multidisciplinary Digital Publishing Institute},
	pages = {2029},
	title = {A {Perceptual} {Evaluation} of {Individual} and {Non}-{Individual} {HRTFs}: A {Case} {Study} of the {SADIE} {II} {Database}},
	volume = {8},
}

@inproceedings{binamix,
  title={Binamix - A Python Library for Generating Binaural Audio Datasets},
  author={Barry, Dan and Shariat Panah, Davoud and Ragano, Alessandro and Skoglund, Jan and Hines, Andrew},
  booktitle={Audio Engineering Society Convention 158},
  year={2025},
  organization={Audio Engineering Society},
url = {https://doi.org/10.48550/arXiv.2505.01369},
}

@article{pulkki1997virtual,
  title={Virtual sound source positioning using vector base amplitude panning},
  author={Pulkki, Ville},
  journal={Journal of the audio engineering society},
  volume={45},
  number={6},
  pages={456--466},
  year={1997},
  publisher={Audio Engineering Society}
}

@article{dietz2011auditory,
  title={Auditory model based direction estimation of concurrent speakers from binaural signals},
  author={Dietz, Mathias and Ewert, Stephan D and Hohmann, Volker},
  journal={Speech Communication},
  volume={53},
  number={5},
  pages={592--605},
  year={2011},
  publisher={Elsevier}
}

@article{soendergaard2013amt,
  title   = {The Auditory Modeling Toolbox},
  author  = {Søndergaard, Pl{\o}ger and Majdak, Piotr},
  journal = {Acta Acustica united with Acustica},
  volume  = {99},
  number  = {6},
  pages   = {988--1002},
  year    = {2013},
  publisher = {S. Hirzel Verlag}
}

@article{scheibler2018pyroomacoustics,
  title   = {Pyroomacoustics: A Python package for audio room simulation and array processing algorithms},
  author  = {Scheibler, Robin and Bezzam, Elie and Dokmani{\'c}, Ivan},
  journal = {Proceedings of the IEEE International Conference on Acoustics, Speech, and Signal Processing (ICASSP)},
  pages   = {351--355},
  year    = {2018}
}

@inproceedings{majdak2013sofa,
  title     = {SOFA: Standardized Open File Format for Acoustics},
  author    = {Majdak, Piotr and Ziegelwanger, Harald and Dietz, Mathias},
  booktitle = {Proceedings of the 134th Audio Engineering Society Convention},
  address   = {Rome, Italy},
  year      = {2013}
}

@inproceedings{yang2021torchaudio,
  title     = {Torchaudio: Building Blocks for Audio and Speech Processing},
  author    = {Yang, Moto Hira and Ni, Zhaoheng and Zhang, Yan and Chuang, Po-Sen Huang and Li, Wei and Shi, Jiatong and Ni, Yifan and others},
  booktitle = {Proceedings of the IEEE International Conference on Acoustics, Speech, and Signal Processing (ICASSP)},
  year      = {2021}
}

@inproceedings{barry2004azimuth,
  title     = {Sound Source Separation: Azimuth Discrimination and Resynthesis},
  author    = {Dan Barry and Bob Lawlor and Eugene Coyle},
  year      = {2004},
  booktitle = {Proceedings of the 7th International Conference on Digital Audio Effects},
  url       = {https://api.semanticscholar.org/CorpusID:2332479}
}

@article{barry2009localization, 
author={Dan Barry and Gavin Kearney}, 
journal={journal of the audio engineering society}, 
title={localization quality assessment in source separation-based upmixing algorithms}, 
year={2009}, 
number={33}, 
month={02},
}

@article{barry2007music, 
author={Dan Barry and Mikel Gainza and Eugene Coyle}, 
journal={journal of the audio engineering society}, 
title={music structure segmentation using the azimugram in conjunction with principal component analysis}, 
year={2007}, 
number={7235}, 
month={10},}

@book{woodworth1954experimental,
  title     = {Experimental Psychology, Revised Edition},
  author    = {Woodworth, Robert S. and Schlosberg, Harold},
  year      = {1954},
  publisher = {Henry Holt},
  address   = {New York, NY}
}

@inproceedings{phokhinanan2023binaural,
  title={Binaural sound localization in noisy environments using frequency-based audio vision transformer (FAViT)},
  author={Phokhinanan, Waradon and Obin, Nicolas and Argentieri, Sylvain},
  booktitle={INTERSPEECH},
  pages={3704--3708},
  year={2023},
  organization={ISCA}
}

@misc{resonanceaudio_matlab,
  title        = {Resonance Audio: MATLAB Tools},
  howpublished = {\url{https://github.com/resonance-audio/resonance-audio/tree/master/matlab}},
  note         = {Accessed: 2025-09-09},
  author       = {{Google Inc.}},
  year         = {2018}
}

@INPROCEEDINGS{birchfield,
  author={Birchfield, S.T. and Gangishetty, R.},
  booktitle={Proceedings. (ICASSP '05). IEEE International Conference on Acoustics, Speech, and Signal Processing, 2005.}, 
  title={Acoustic localization by interaural level difference}, 
  year={2005},
  volume={4},
  number={},
  pages={iv/1109-iv/1112 Vol. 4},
  doi={10.1109/ICASSP.2005.1416207}}

@article{kuang2025bast,
  title={BAST-Mamba: Binaural Audio Spectrogram Mamba Transformer for binaural sound localization},
  author={Kuang, Sheng and Shi, Jie and van der Heijden, Kiki and Mehrkanoon, Siamak},
  journal={Neurocomputing},
  pages={130804},
  year={2025},
  publisher={Elsevier}
}

@INPROCEEDINGS{rickard2000duet,
  author={Jourjine, A. and Rickard, S. and Yilmaz, O.},
  booktitle={2000 IEEE International Conference on Acoustics, Speech, and Signal Processing. Proceedings (Cat. No.00CH37100)}, 
  title={Blind separation of disjoint orthogonal signals: demixing N sources from 2 mixtures}, 
  year={2000},
  volume={5},
  number={},
  pages={2985-2988 vol.5},
  doi={10.1109/ICASSP.2000.861162}}

@article{DolbyAtmos,
title = "Practitioners' Perspectives on Spatial Audio: Insights into Dolby Atmos and Binaural Mixes in Popular Music",
author = "Christopher Dewey and Austin Moore and Hyunkook Lee",
year = "2024",
month = jul,
day = "9",
language = "English",
volume = "72",
pages = "504--516",
journal = "AES: Journal of the Audio Engineering Society",
issn = "0004-7554",
publisher = "Audio Engineering Society",
number = "7/8",
}

@misc{iamf,
  author       = {{Alliance for Open Media}},
  title        = {Immersive Audio Metadata Format (IAMF)},
  year         = {2025},
  howpublished = {\url{https://aomediacodec.github.io/iamf/}},
  note         = {Accessed: 2025-01-28}
}

@article{rane2022survey,
  author       = {M. Rane and P. Coleman and R. Mason and S. Bech},
  title        = {Survey of User Perspectives on Headphone Technology},
  journal      = {Journal of the Audio Engineering Society},
  year         = {2022},
  month        = {05},
  number       = {10556},
  url          = {https://aes2.org/publications/elibrary-page/?id=21669}
}

@article{ambiqual,
  author       = {M. Narbutt and J. Skoglund and A. Allen and M. Chinen and D. Barry and A. Hines},
  title        = {AMBIQUAL: Towards a Quality Metric for Headphone Rendered Compressed Ambisonic Spatial Audio},
  journal      = {Applied Sciences},
  year         = {2020},
  volume       = {10},
  number       = {9},
  pages        = {3188},
  doi          = {10.3390/app10093188},
  url          = {https://doi.org/10.3390/app10093188}
}

@misc{Binamix2025,
  author       = {Dan Barry and Davoud Shariat Panah and Alessandro Ragano and Jan Skoglund and Andrew Hines},
  title        = {Binamix: A Python Library for Generating Binaural Audio Datasets},
  year         = {2025},
  url          = {https://github.com/QxLabIreland/Binamix/},
  note         = {Accessed: 2025-04-01}
}

@inproceedings{rudzki2025eclipsa,
  title     = {On the Design of the Binaural Rendering Library for Eclipsa Audio Immersive Audio Container},
  author    = {Tomasz Rudzki and Gavin Kearney and Jan Skoglund},
  year      = {2025},
  booktitle = {Proceedings AES 158th Audio Engineering Society Convention (2025)}
}

@article{panah2025binaqual,
  title={BINAQUAL: A Full-Reference Objective Localization Similarity Metric for Binaural Audio},
  author={Panah, Davoud Shariat and Barry, Dan and Ragano, Alessandro and Skoglund, Jan and Hines, Andrew},
  journal={arXiv preprint arXiv:2505.11915},
  year={2025}
}

\end{document}